\documentclass[nofootinbib,twocolumn,superscriptaddress,longbibliography,amsmath,amssymb,aps,prl]{revtex4-1} 
\usepackage{latexsym}
\usepackage{graphics}
\usepackage{bm}
\usepackage[normalem]{ulem}
\usepackage{graphicx} 
\usepackage{color}
\usepackage{amsmath}
\usepackage{amssymb}
\usepackage{feynmp-auto}
\usepackage{slashed}
\usepackage{enumerate}
\usepackage{amsfonts, amssymb,amsmath,latexsym,graphics, graphicx,epsfig,multirow,comment,
,feyn,slashed,xcolor,afterpage, makecell} 
\usepackage{booktabs, blindtext}
\usepackage{tabularx,afterpage}
\usepackage{soul}
\usepackage[colorlinks=true
,urlcolor=blue
,anchorcolor=blue
,citecolor=blue
,filecolor=blue
,linkcolor=blue
,menucolor=blue
,linktocpage=true
,pdfproducer=medialab
,pdfa=true
]{hyperref}
\usepackage[utf8]{inputenc}
\usepackage{url}

\newcolumntype{L}[1]{>{\raggedright\let\newline\\\arraybackslash\hspace{0pt}}m{#1}}
\newcolumntype{C}[1]{>{\centering\let\newline\\\arraybackslash\hspace{0pt}}m{#1}}
\newcolumntype{R}[1]{>{\raggedleft\let\newline\\\arraybackslash\hspace{0pt}}m{#1}}

\newcommand {\be} {\begin {equation}}
\newcommand {\ee} {\end {equation}} 
\newcommand {\nn} {\nonumber}
\newcommand {\bes} {\begin {equation*}}
\newcommand {\ees} {\end {equation*}}

\newcommand{\beq}{\begin{equation}}
\newcommand{\eeq}{\end{equation}}
\newcommand{\bea}{\begin{eqnarray}}
\newcommand{\eea}{\end{eqnarray}}

\newcommand{\Ad}{{\gamma_d}}
\newcommand{\hd}{{h_d}}
\newcommand{\nAd}{{n_{\gamma_d}}}
\newcommand{\nhd}{{n_{h_d}}}
\newcommand{\nAdsq}{n_{\gamma_d}^2}
\newcommand{\nhdsq}{n_{h_d}^2}

\newcommand{\nhdcu}{n_{h_d}^3}
\newcommand{\nAdeq}{n_{\gamma_d}^{\rm eq}}
\newcommand{\nhdeq}{n_{h_d}^{\rm eq}}
\newcommand{\Td}{{T_d}}
\newcommand{\TSM}{{T_\text{SM}}}

\newcommand{\mAd}{{m_{\gamma_d}}}
\newcommand{\mhd}{{m_{h_d}}}

\newcommand{\eqt}{{\rm eq}}
\newcommand{\lra}{\leftrightarrow}
\newcommand{\ra}{\rightarrow}
\newcommand{\psid}{{\psi}}

\newcommand{\nAeq}{n_{A}^{\rm eq}}
\newcommand{\nBeq}{n_{B}^{\rm eq}}
\newcommand{\xG}{x^\Gamma}
\newcommand{\RG}{R_\Gamma}

\usepackage{csvsimple}

\begin{document}

\title{Dark Higgs Dark Matter}

\preprint{PUPT XXXX}

\author{Cristina Mondino}
\affiliation{Center for Cosmology and Particle Physics, Department of Physics, New York University, New York, NY 10003, USA}

\author{Maxim Pospelov}
\affiliation{William I. Fine Theoretical Physics Institute, School of Physics and Astronomy, University of Minnesota, Minneapolis, MN 55455, USA}

\author{Joshua T. Ruderman}
\affiliation{Center for Cosmology and Particle Physics, Department of Physics, New York University, New York, NY 10003, USA}

\author{Oren Slone}
\affiliation{Princeton Center for Theoretical Science, Princeton University, Princeton, NJ 08544, USA}

\date{\today}

\begin{abstract}
A new $U(1)$ ``dark" gauge group coupled to the Standard Model (SM) via the kinetic mixing portal provides a natural dark matter candidate in the
form of the Higgs field, $h_d$, responsible for generating the mass of the dark photon, $\gamma_d$. We show that the condition $m_{h_d}\leq m_{\gamma_d}$, 
together with smallness of the kinetic mixing parameter, $\epsilon$, and/or dark gauge coupling, $g_d$, leads the dark Higgs to be sufficiently metastable to constitute dark matter. We analyze the Universe's thermal history and show that both freeze-in, ${\rm SM}\to \{\gamma_d, h_d\}$,  and freeze-out, $ \{\gamma_d, h_d\} \to {\rm SM}$, processes can lead to viable dark Higgs dark matter with a sub-GeV mass and a kinetic mixing parameter in the range $10^{-13}\lesssim\epsilon\lesssim10^{-6}$.  Observable signals in astrophysics and cosmology include modifications to primordial elemental abundances, altered energetics of supernovae explosions, dark Higgs decays in the late Universe, and dark matter self-interactions.

\end{abstract}
\maketitle

\emph{Introduction.} Evidence for dark matter constitutes one of the strongest arguments for extending the Standard Model (SM) with {\em Dark Sector(s)} (DS)\@. One of the simplest examples is an additional $U(1)_d$ gauge symmetry associated with its gauge boson, the dark photon $\Ad$, that mediates DS-SM interactions. It is commonly assumed that Dark Matter (DM) is charged under  $U(1)_d$ and is represented by additional states in the DS\@.

A great deal of theoretical and experimental attention has been devoted to the study of such a DS \cite{Battaglieri:2017aum,Beacham:2019nyx} in recognition of the fact that its mass scale can be at, or below 1\,GeV, offering a variety of new probes. In this study, we show that an even {\em more minimal} option exists: the field responsible for generating the mass of $\Ad$, a dark Higgs $\hd$, is a viable DM candidate with different phenomenological implications than those commonly assumed~\cite{Battaglieri:2017aum,Beacham:2019nyx}. Throughout the paper, we analyze the salient features of Dark Higgs Dark Matter (DHDM), including its (meta)-stability and genesis through cosmic history.

For such a model, the relevant DS Lagrangian is
\bea
\mathcal{L_{\rm DS}} & = & |D^\mu\phi|^2 - \frac{1}{4} (F_d^{\mu \nu})^2 -\frac{\epsilon}{2} F_d^{\mu\nu}F_{\mu\nu} - V(\phi)  \, ,
\eea
where $\phi$ is a charged scalar field, $F^{\mu\nu}$ and $F_{d}^{\mu\nu}$ are the SM and dark photon field strengths, and $\epsilon$ is the kinetic mixing parameter. The scalar potential, $V(\phi) = -\mu^2 |\phi|^2 + \lambda |\phi|^4$, generates a non-zero Vacuum Expectation Value (VEV) for $\phi$, $\langle \phi \rangle = \mu/\sqrt{2\lambda} \equiv v/\sqrt{2}$, that spontaneously breaks the $U(1)_d$ gauge symmetry. Expanding $\phi$ around the VEV, $\phi = (v+\hd)/\sqrt{2}$, gives the mass terms for the dark Higgs, $\mhd = \sqrt{2\lambda} v$, and for the dark photon, $\mAd = g_d v$, where $g_d$ is the dark gauge coupling, and one can define $\alpha_d \equiv g_d^2/4\pi$.

For a natural choice of couplings, $\lambda \sim g_d^2$, the dark particle's masses are expected to be of similar order, $\mhd \sim \mAd$. In particular, it is possible for $\hd$ to have a mass below, or close to that of $\Ad$, with a potentially long lifetime~\cite{Batell:2009yf,Darme:2017glc}.

\begin{figure}
  \centering
  \includegraphics[width=3.4in]{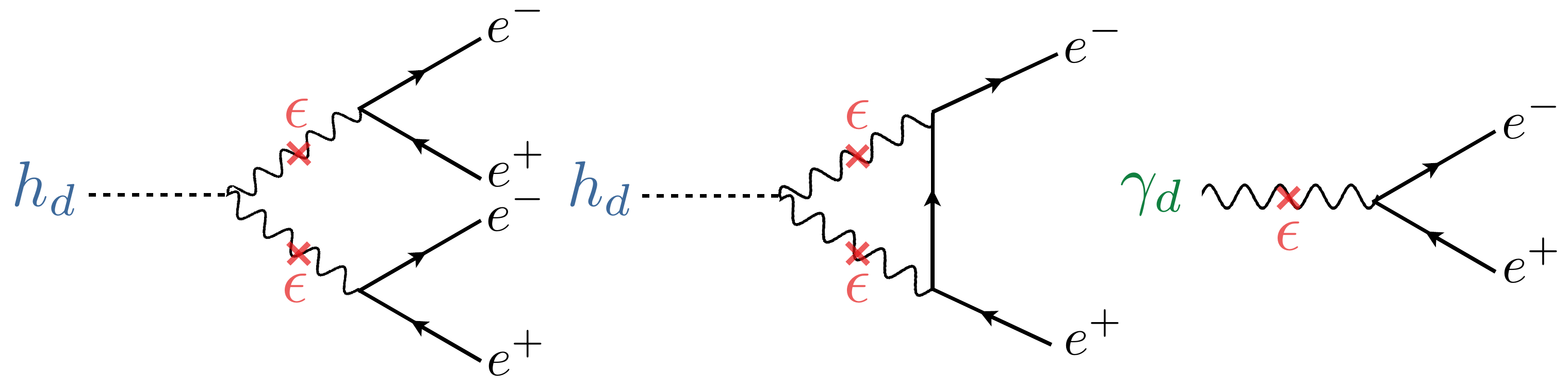}
  \caption{Decay processes of dark sector particles. Left and central diagrams are dark Higgs decay channels available for $4m_e < \mhd < 2m_\mu$ and $\mhd <\mAd$. The right diagram is the dark photon decay channel for $2m_e < \mAd < 2m_\mu$.}
  \label{fig:DarkHiggs_Decay}
\end{figure}

When $\mhd < \mAd$, the only open decay channels for an MeV-scale dark Higgs are $\hd \rightarrow e^+e^-e^+e^-$, and the loop-induced process, $\hd \rightarrow e^+e^-$. The associated Feynman diagrams are depicted in Fig.~\ref{fig:DarkHiggs_Decay} and require double insertions of $\epsilon$, leading to a dark Higgs lifetime that scales as $\tau_\hd \propto \epsilon^{-4}$ (for full expressions of the decay widths see {\it e.g.}~ Ref.~\cite{Batell:2009yf}). An additional Higgs portal coupling could lead to SM-DS Higgs mixing and further decay channels for $h_d$. However, if such a coupling is absent at tree level, it only appears at one loop, and the resulting SM-DS Higgs mixing parameter 
$\theta_{h-h_d} \propto (\epsilon g_d)^2 (v/v_{\rm EW})$ is negligibly small.

Combining the decay channels for $4m_e \ll \mhd < 2m_\mu$, the approximate decay width for DHDM is
\begin{eqnarray}
\tau_U\Gamma_{2e,4e} \simeq 8\times 10^{-8} \left[\frac{\epsilon}{10^{-9}}\right]^4 \left[\frac{\alpha_d}{10^{-4}}\right]  \left[\frac{m_{h_d}}{100\,{\rm MeV}}\right]f. \, 
\end{eqnarray}
Here, $\tau_U$ is the age of the Universe, and $f=f(m_{h_d}/m_{\gamma_d})$ is a dimensionless function of order unity at $m_{h_d}= m_{\gamma_d}$. Note that $\tau_U\Gamma_{2e,4e}<1$ is an insufficient condition for $h_d$ to be DM, since stronger bounds are imposed by limits on diffuse photon spectra~\cite{Essig:2013goa} and precision measurements of CMB anisotropies~\cite{Slatyer:2016qyl,Poulin:2016anj}.   

\begin{figure}
  \centering
   \includegraphics[width=2.5in]{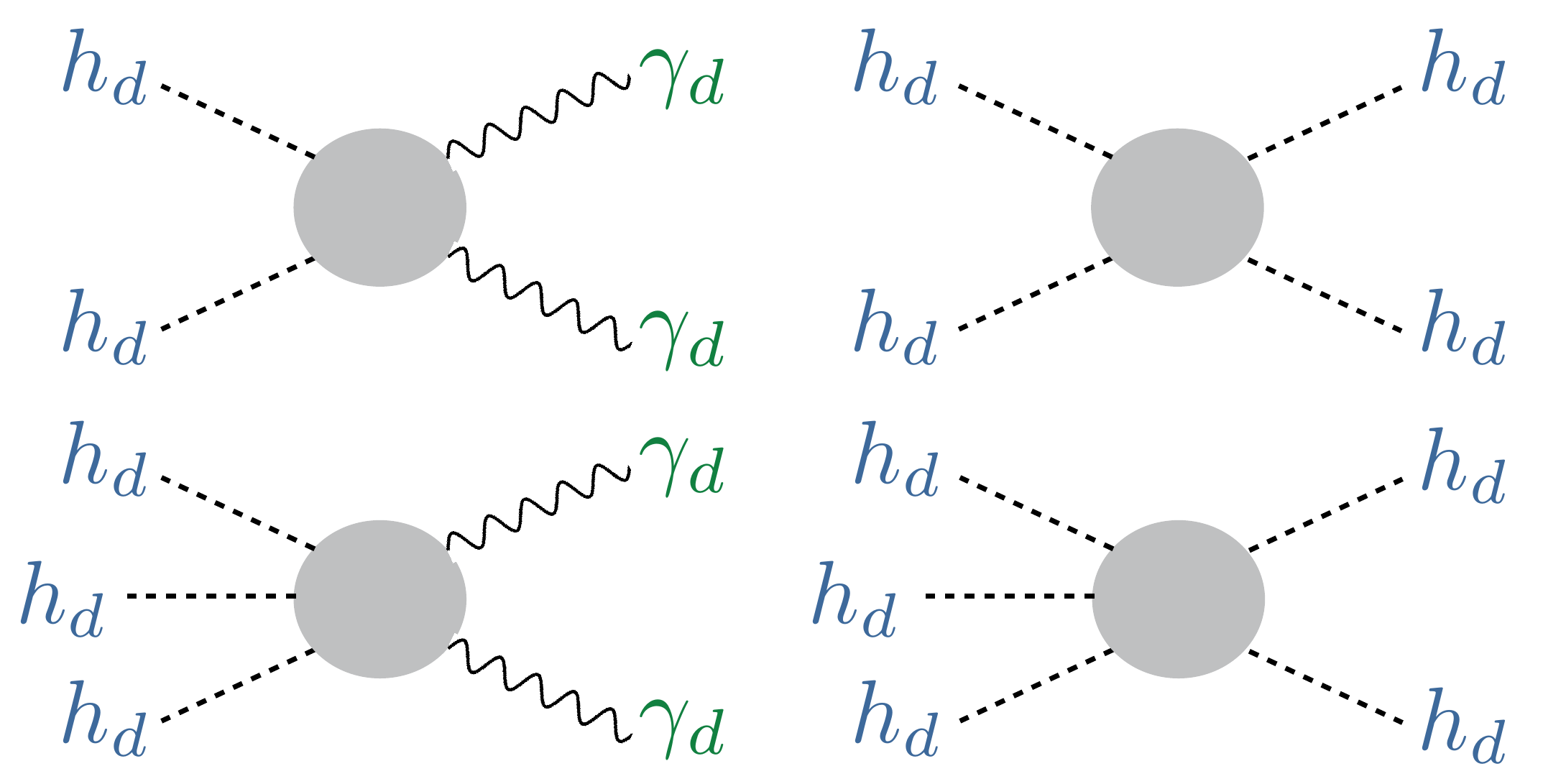}
   \caption{Diagrams for processes involving dark sector particles only. Rates of $2\leftrightarrow 2$ processes on the top row scale as $\alpha_d^2$ while $3\leftrightarrow 2$ processes on the bottom row scale as $\alpha_d^3$.}
  \label{fig:BE_Diagrams}
\end{figure}

\emph{Brief overview of DHDM cosmic history.} 
Various processes affect the DS's thermal history. Those that exchange energy between the two sectors have rates that scale as $\epsilon^2$. Of these, dark photon decay and inverse decay, $\Ad \lra e^+e^-$, (shown in Fig.~\ref{fig:DarkHiggs_Decay}) always dominate over the scattering processes $\Ad e^\pm \lra \gamma e^\pm$, $\hd\Ad \lra e^+e^-$, and $\hd e^\pm \lra \Ad e^\pm$, in this study. Interactions involving only DS particles are controlled by $\alpha_d$. Both $2 \leftrightarrow 2$ and $3 \leftrightarrow 2$ processes (shown in Fig.~\ref{fig:BE_Diagrams}) are relevant, with rates that scale as $\alpha_d^2$ and $\alpha_d^3$, respectively. Given the above ingredients, the DS is described by the parameters $\{\mAd,\,\mhd,\,\alpha_d,\,\epsilon\}$, which control both the lifetime and abundance of $h_d$.

\begin{figure}
  \centering
  \includegraphics[width=0.45\textwidth]{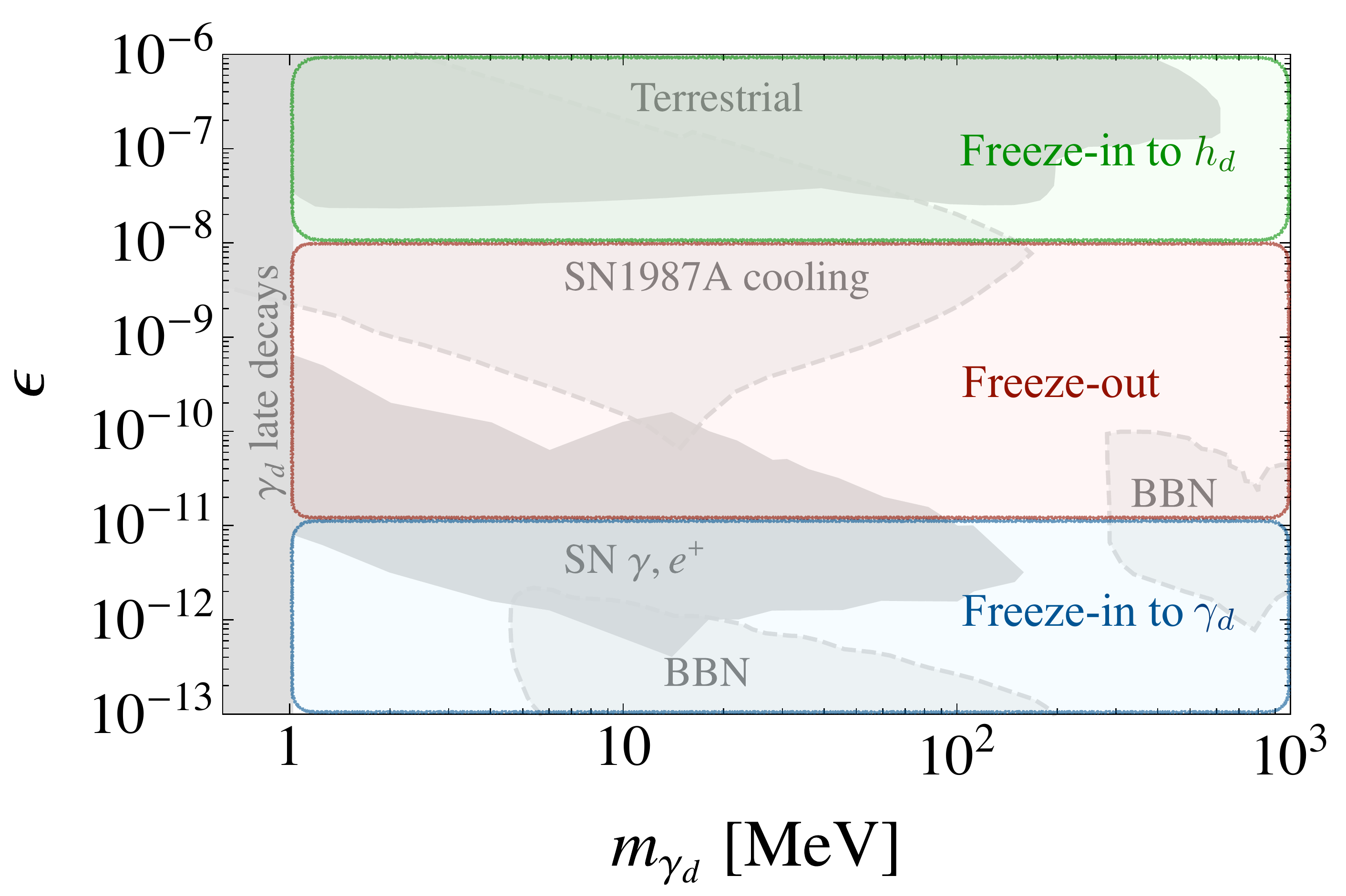}
   \caption{Illustration of various DHDM production regimes considered in this study: freeze-in to $\Ad$ followed by annihilation to $\hd$ at tiny kinetic mixing, freeze-out of secluded annihilation at intermediate values, and direct freeze-in to $\hd$ with $\Ad$ in equilibrium with the SM at large $\epsilon$. Note that the true boundaries between the different regimes are $\mAd$ and $\mhd$ dependent. Gray shaded regions show existing bounds on the dark photon, reproduced from Refs.~\cite{Fradette:2014sza, Chang:2016ntp, DeRocco:2019njg}. Lightly shaded regions are potentially modified by the presence of the dark Higgs field, with an $\alpha_d$ dependence.}
   \label{fig:mgeps_cartoon}
\end{figure}

This study identifies three regimes of interest, schematically represented in Fig.~\ref{fig:mgeps_cartoon} and corresponding to different production mechanisms and phenomenology.
When the kinetic mixing is tiny, the DS is thermally decoupled from the SM bath and production of DHDM proceeds via freeze-in of the dark photon particles, which later annihilate into the dark Higgs. This is allowed for several decades of $\alpha_d$ and $\mAd$, when $\epsilon \sim 10^{-13} - 10^{-11}$.
Alternatively, the DS can be independently populated during reheating, reaching internal thermal equilibrium as long as $\alpha_d$ is sufficiently large. For an intermediate range of kinetic mixing, $\epsilon \sim 10^{-11} - 10^{-8}$, the DS never fully thermalizes with the SM\@. However, the dark Higgs can deplete its number density by annihilating into dark photons that eventually decay into SM particles. 
Again, the correct DHDM abundance can be obtained for several decades of $\alpha_d$ and $\mAd$. 
Finally, going to $\epsilon \gtrsim 10^{-8}$ requires increasing $\tau_\hd$ by choosing a very small $\alpha_d$. In this case, the dark photon reaches thermal equilibrium with the SM and the dark Higgs can freeze-in from the thermal bath of $\Ad$. An upper limit of $\epsilon \lesssim 10^{-6}$ follows from constraints on $\tau_\hd$.

Assuming a thermal distribution for the DS particles, the above regimes can be described with a set of three coupled Boltzmann Equations (BEs). The first two, $0$-th moment BEs, track the number densities of each species,
\bea
\dot{n}_{\hd} + 3Hn_{\hd} & = & \mathcal{C}_{2\Ad \ra 2\hd} - \mathcal{C}_{3\hd \ra 2\Ad}\label{eq:boltz_nhd} \, , \\
\dot{n}_{\Ad} + 3Hn_{\Ad} & = & -\mathcal{C}_{\Ad \ra e^+e^-} - \mathcal{C}_{2\Ad \ra 2\hd} + \frac{2}{3} \mathcal{C}_{3\hd \ra 2\Ad}, \,\,\,\,\,\, \label{eq:boltz_nAd} \,
\eea
where $H$ is the Hubble expansion rate and the collision terms, $\mathcal{C}_{{\rm init} \to {\rm fin}}$, correspond to the processes of Fig.~\ref{fig:BE_Diagrams} and the $\Ad$ decay of Fig.~\ref{fig:DarkHiggs_Decay} (the explicit expressions appear in Appendix~\ref{sec:BE_terms}). The third, $1$-st moment BE, tracks the DS temperature $T_d$. In the nonrelativistic limit, when $\Td \lesssim \mhd, \mAd$, 
\bea
n \frac{\dot{T_d}}{T_d} + 2 H n & = & \mathcal{C}^{(E)}_{\Ad \ra e^+e^-} + \mathcal{C}^{(E)}_{2\Ad \ra 2\hd} + \frac{2}{3}\mathcal{C}^{(E)}_{3\hd \ra 2\Ad} \, , \,\,\,\,\,\,\,\,
\label{eq:boltz_Td}
\eea
where $n \equiv \nAd + \nhd$. In Eqs.~(\ref{eq:boltz_nhd}-\ref{eq:boltz_Td}), only dominant processes have been included.

\begin{figure}
  \centering
  \includegraphics[width=0.45\textwidth]{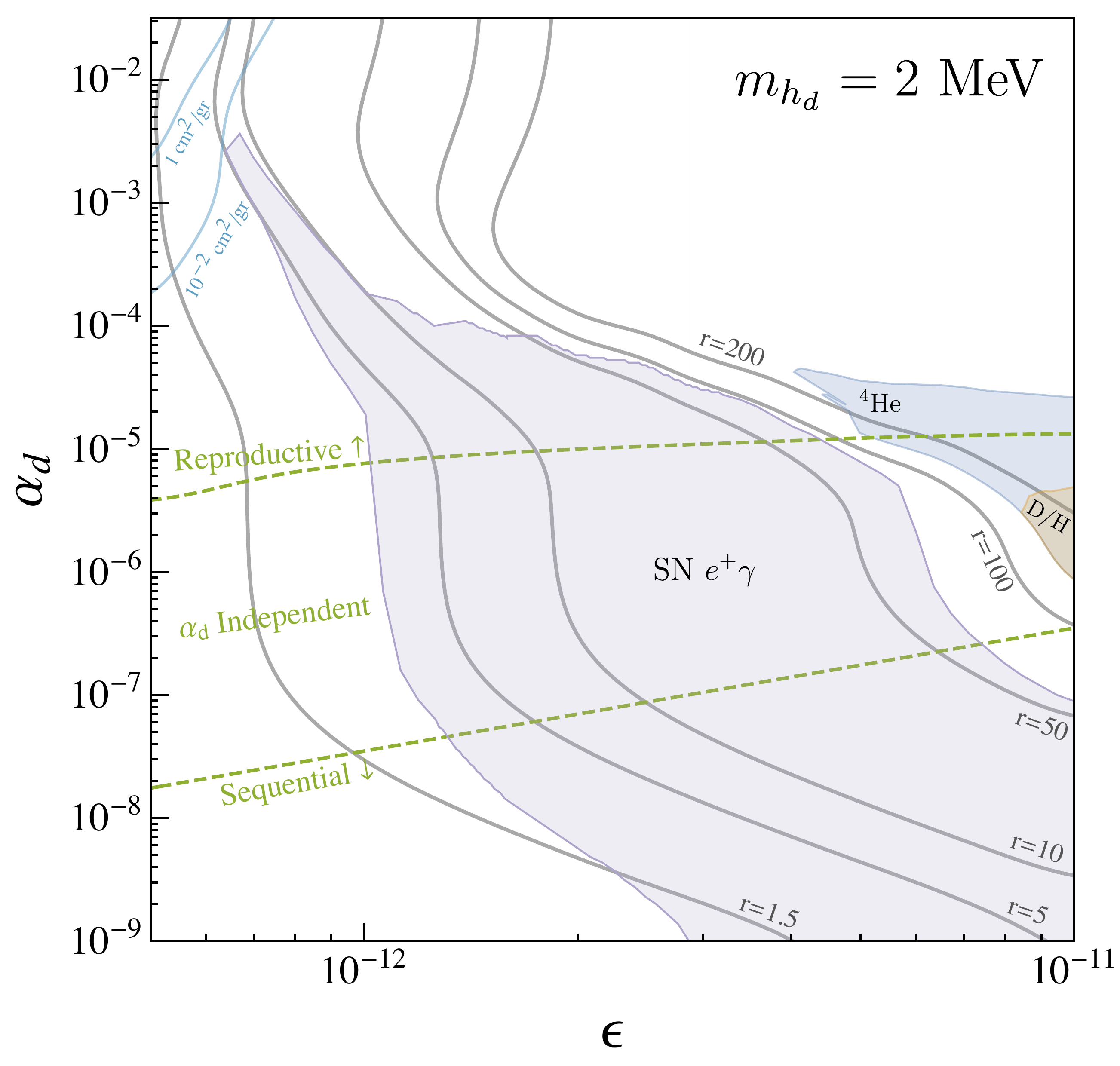}
    \caption{Parameter space for the freeze-in to dark photon regime for $\mhd = 2$ MeV\@. The correct DHDM relic density is produced everywhere by fixing $r \equiv \mAd/\mhd$; constant values are shown as gray contours. Three qualitatively different regions are evident: sequential freeze-in where only a fraction of $\Ad$ annihilate into $\hd$, an $\alpha_d$ independent regime where essentially all $\Ad$ annihilate into $\hd$, and a reproductive regime where $2\to3$ processes become important. Also shown are bounds from the non-observation of $e^+$ and $\gamma$ signals of $\Ad$ production within SNe~\cite{DeRocco:2019njg}, and BBN constraints from overproduction of $^4$He and D~\cite{Berger:2016vxi}. Blue contours correspond to constant self-interaction cross sections per unit mass, $\sigma_{\rm SIDM}/\mhd$.}
  \label{fig:FI_alpha_eps}
\end{figure}

\emph{Freeze-In Regimes.} At extremely small $\epsilon$, the DS is very weakly coupled to the SM\@. Assuming a negligible production of DS particles during reheating, {\it i.e.~}setting the initial $\nhd$ and $\nAd$ to zero in the BEs (\ref{eq:boltz_nhd}-\ref{eq:boltz_Td}), dark photons can freeze-in from SM particles via out-of-equilibrium processes~\cite{Pospelov:2008jk,Redondo:2008ec,Hall:2009bx,Fradette:2014sza}, dominated by the inverse decay diagram shown on the right of Fig.~\ref{fig:DarkHiggs_Decay}. After a population of $\Ad$ develops, these annihilate into dark Higgs particles through $\Ad\Ad \to \hd\hd$, since $\mhd < \mAd$. 
The result of such a freeze-in to $\Ad$ scenario is shown in Fig.~\ref{fig:FI_alpha_eps}, for the choice $\mhd = 2$\,MeV\@. At every point, 
the ratio $r \equiv \mAd/\mhd$ is chosen to match the observed DM relic density, $\Omega_\hd h^2 = 0.12$~\cite{Aghanim:2018eyx}. In various regions, multiple $r$ values satisfy this condition; the lowest has been chosen everywhere. Gray contours correspond to constant values of $r$. Three qualitatively different regimes occur, delineated by dashed green lines.

Two of these regimes occur when $\alpha_d$ is small enough such that $2 \lra 3$ processes are negligible. Then, the total DS number density, $n$, can be calculated by summing Eqs.~\eqref{eq:boltz_nhd} and~\eqref{eq:boltz_nAd}. This results in the following yield, $Y \equiv n/s$ ($s$ is the entropy density of the Universe), at late times,
\beq
Y \approx \frac{3 m^3_\Ad \Gamma^0_{\Ad} }{2\pi^2 r} \int_{x \ll 1}^{x \approx 10} \frac{K_1(r x)}{x^2 H s} dx \approx 7 \cdot 10^{-5} \frac{\epsilon^2 m_{\rm pl}}{\mAd} \, ,
\label{eq:FI_YAd}
\eeq
where $x \equiv \mhd/\TSM$, $\TSM$ is the SM temperature, $K_1$ is the modified Bessel function, $m_{\rm pl}$ is the Planck mass and to a good approximation one can use the decay width $\Gamma^0_{\Ad} \approx \alpha \epsilon^2 \mAd/3$ with $\alpha$ the SM fine structure constant. In the above equation we have taken $g_{*} \approx g_{*s} \approx 10$. The final yield of $\hd$ depends on the efficiency of the annihilation process, $\Ad \Ad \to \hd \hd$, which is controlled by $r$ and $\alpha_d$. 

For a given mass ratio, when $\alpha_d$ is small enough such that most dark photons do not annihilate, the result can be approximated by taking $Y \to Y_{\Ad}$ in Eq.~\eqref{eq:FI_YAd}, and plugging into Eq.~\eqref{eq:boltz_nhd} while setting $\mathcal{C}_{3\hd \ra 2\Ad}=0$. Then, the rate for generating DHDM from $\gamma_d$ scales as $\propto Y_{\Ad}^2\alpha^2_dm^{-2}_\Ad$. Neglecting inverse annihilation, one finds
\beq
\frac{\Omega_\hd h^2}{0.12} \approx \left[\frac{\epsilon}{5\cdot10^{-12}}\right]^4 \left[\frac{\alpha_d}{10^{-9}}\right]^2 \left[\frac{\mhd}{2\text{ MeV}}\right] \left[\frac{4\text{ MeV}}{\mAd}\right]^3.
\label{eq:FI_regime_1}
\eeq
Thus, the $\hd$ relic density is fixed along contours of constant $\epsilon^4 \alpha_d^2$. This is a realization of the sequential freeze-in mechanism, identified for a different model by Ref.~\cite{Hambye:2019dwd}.

For larger $\alpha_d$ values, $\Ad \Ad \to \hd \hd$ annihilations become extremely efficient and essentially all dark photons convert into dark Higgs particles. The final DHDM abundance can then be approximated by replacing $Y \to Y_{\hd}$ in Eq.~\eqref{eq:FI_YAd}. The result is now $\alpha_d$ independent,
\beq
\frac{\Omega_\hd h^2}{0.12} \approx \left[\frac{\epsilon}{10^{-12}}\right]^2 \left[\frac{\mhd}{2 \text{ MeV}}\right] \left[\frac{4 \text{ MeV}}{\mAd}\right].
\label{eq:FI_regime_2}
\eeq

\begin{figure}
  \centering
  \includegraphics[width=0.419\textwidth]{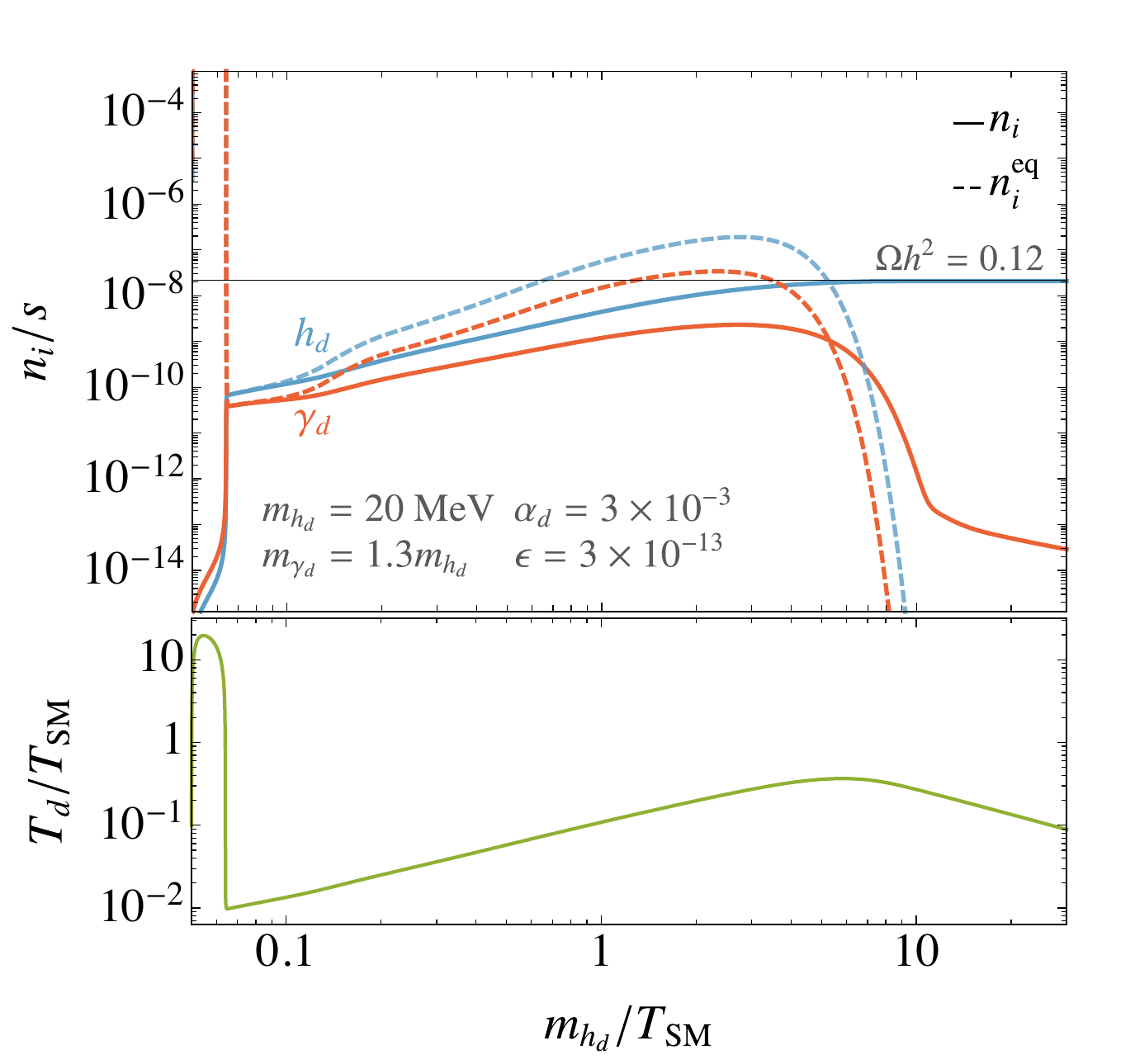}
  \caption{Example of the DS evolution in the reproductive freeze-in regime. The top panel shows the evolution of the $\hd$ and $\Ad$ comoving number densities. The bottom panel shows the temperature ratio, $\Td/\TSM$. Dashed curves correspond to equilibrium distributions.}
  \label{fig:FreezeIn_1d}
\end{figure}

The third regime occurs when $\alpha_d$ becomes large enough  that number changing processes within the DS are important. A particularly interesting phenomenon occurs when the rate for $\Ad\Ad \lra \hd\hd$ is faster than the Hubble rate, and $2 \ra 3$ processes are simultaneously active. The former imposes chemical equilibrium between $\hd$ and $\Ad$, whereas the latter pushes their initially negative chemical potentials to zero. This combination causes a rapid drop in $\Td$ via removal of kinetic energy, setting $\nAd \approx \nAdeq$ and $\nhd \approx \nhdeq$ (number densities with zero chemical potentials), washing out any dependence on the dark temperature's initial condition. We denote this production mechanism {\it reproductive freeze-in}, and provide further details in Appendix~\ref{sec:ReproductiveDM}. A hidden sector with thermodynamics driven by $3 \ra 2$ processes undergoes a phase known as ``cannibalism"~\cite{Carlson:1992fn}, while here $2 \ra 3$ processes are essential; in this sense reproduction is the opposite of cannibalism. An example of numerical solutions for $\nAd$, $\nhd$, and $T_d$, in the reproductive regime, is shown in Fig.~\ref{fig:FreezeIn_1d}. The sharp drop in $T_d$, which proceeds until $\nAd\approx\nAdeq$ and $\nhd\approx\nhdeq$, is evident. A large number of dark particles are produced from the $2 \to 3$ processes. Thus, the correct relic density is achieved for smaller values of $\alpha_d$, as can be seen by the behavior of the gray contours in Fig.~\ref{fig:FI_alpha_eps}.

As discussed above, freeze-in can also occur for $\epsilon \gtrsim 10^{-8}$, when dark photons are in thermal equilibrium with the SM\@. For tiny dark gauge couplings, dark Higgs particles freeze-in from the SM and $\Ad$ plasma, through the processes $\Ad\Ad \to \hd\hd$, $e^+e^- \to \Ad\hd$, and $e\Ad \to e\hd$ (the relative contributions of these depend on $\epsilon$, $\mAd$, and $\mhd$). The DM density is obtained approximately with $\alpha_d\sim10^{-13}$. A similar scenario for freeze-in of the Higgs field of a $U(1)_{B-L}$ symmetry has recently been considered in Ref.~\cite{Mohapatra:2020bze}.

\emph{Phenomenological Consequences.} The phenomenology of the freeze-in to dark photon scenario is summarized in Fig.~\ref{fig:FI_alpha_eps}. Dark photons produced within supernovae (SNe) can decay outside the explosion and produce positron annihilation or gamma-ray signals~\cite{DeRocco:2019njg}. Non-observation of these signals from galactic or extragalactic SNe exclude a region in the \{$\mAd,\epsilon$\} parameter space which has been converted into the purple region in the figure. 

Additional bounds come from the effects of late dark photon decays on primordial nucleosynthesis~\cite{Fradette:2014sza,Berger:2016vxi}. When $\Ad$ is heavier than two pions, decays into hadrons alter the $n/p$ ratio and destroy light elements, such as deuterium and helium. Measurements of D/H and $^4$He exclude the brown and the blue regions. Note that these constraints are inferred from Ref.~\cite{Berger:2016vxi}, imposing that $\mAd$ is above the pion decay threshold.

Finally, DHDM can have sizable self interactions, $\hd\hd \lra \hd\hd$, with a cross section, $\sigma_{\rm SIDM}/\mhd = 18 \pi \alpha_d^2 \mhd /m^4_\Ad$. The blue contours correspond to constant $\sigma_{\rm SIDM}/\mhd$. In the upper left corner values of 1 cm$^2$/gr can be achieved; of interest for astrophysical signals of DM~\cite{Spergel:1999mh}.

\emph{Freeze-Out.} For intermediate values of $\epsilon$, the DHDM relic density can be set by the freeze-out of secluded annihilations in the early Universe. Interestingly, the freeze-out scenario of this work combines many properties which have previously been studied in Refs.~\cite{Carlson:1992fn, Pappadopulo:2016pkp, Farina:2016llk, Dror:2016rxc, Griest:1990kh, DAgnolo:2015ujb, Cline:2017tka}. The salient features of this regime can be summarized as follows. The DS is initially decoupled from the SM with $\Td \ne \TSM$, and has a mass gap and a small hierarchy 
with $\mhd \lesssim \mAd$. The processes described in Fig.~\ref{fig:BE_Diagrams} 
provide active number changing interactions which lead to a period of DS self-depletion and self-heating. Additional exponential depletion of DS particles occurs through out-of-equilibrium decay of $\Ad$, leading to a delayed freeze-out of $\hd\hd \to \Ad\Ad$, with a non-zero dark chemical potential.

The detailed evolutions of $\nhd, \nAd$, and $\Td$ follow from the same set of BEs~(\ref{eq:boltz_nhd}-\ref{eq:boltz_Td}) as before.
Initial conditions are set by assuming that the DS is independently populated during reheating and subsequently reaches internal thermal equilibrium. At high temperatures, the SM and DS evolve as two decoupled baths and, for small $\epsilon$, they never equilibrate. Indeed, energy exchanging processes between the two sectors, dominated by $\Ad \lra e^+e^-$ at $\TSM \approx \mAd \approx \mhd$, remain slower than the Hubble rate for $\epsilon \lesssim 10^{-8} (\mAd/100\text{ MeV})^{1/2}$. However, even below this threshold, the small coupling between sectors can still affect the DS's thermal history. For larger values of $\epsilon$, energy exchange becomes efficient, 
leading to the standard secluded freeze-out scenario with $\Td = \TSM$~\cite{Evans:2017kti}.

In the decoupled limit, if number changing processes remain active when $\Td < \text{min}[\mAd,\mhd]$, the DS maintains vanishing chemical potentials and undergoes a period of cannibalism, during which $\Td$ decreases logarithmically with the scale factor~\cite{Carlson:1992fn, Pappadopulo:2016pkp}. This slow cooling occurs because rapid $3 \rightarrow 2$ processes produce kinetic energy. 
Cannibalism proceeds as long as $(n^\text{eq}_\hd)^2 \langle \sigma v^2 \rangle_{\hd\hd\hd \ra \Ad\Ad} \gtrsim H$ at $\Td \approx \TSM \approx \mhd$, which corresponds to $\alpha_d \gtrsim 10^{-6} (\mhd/\text{MeV})^{1/3}$.

The freeze-out itself proceeds via the dominant annihilation channel, $\hd\hd \rightarrow \Ad\Ad$, which efficiently removes DHDM particles only when the decay process $\Ad \to e^+e^-$ becomes active~\cite{Farina:2016llk, Dror:2016rxc}. The latter also cools down the DS, inducing the approximate scaling $\Td \propto T^2_\text{SM}$ (typical of a nonrelativistic decoupled sector). Since $\hd\hd \rightarrow \Ad\Ad$ is kinematically forbidden at zero temperature, the annihilation rate becomes exponentially suppressed when $\Td \lesssim (\mAd-\mhd)$. The kinematic suppression here is even stronger than in the original forbidden DM scenario~\cite{Griest:1990kh, DAgnolo:2015ujb}, due to the faster cooling of the DS with respect to the SM bath. Therefore, the final DHDM relic density is exponentially sensitive to the mass splitting and to the temperature at which $\Ad$ decays become active (the latter determines the onset of the scaling $\Td \propto T^2_\text{SM}$). Further details are provided in Appendix~\ref{sec:FO_analytic}. 

\begin{figure}
  \centering
  \includegraphics[width=0.4\textwidth]{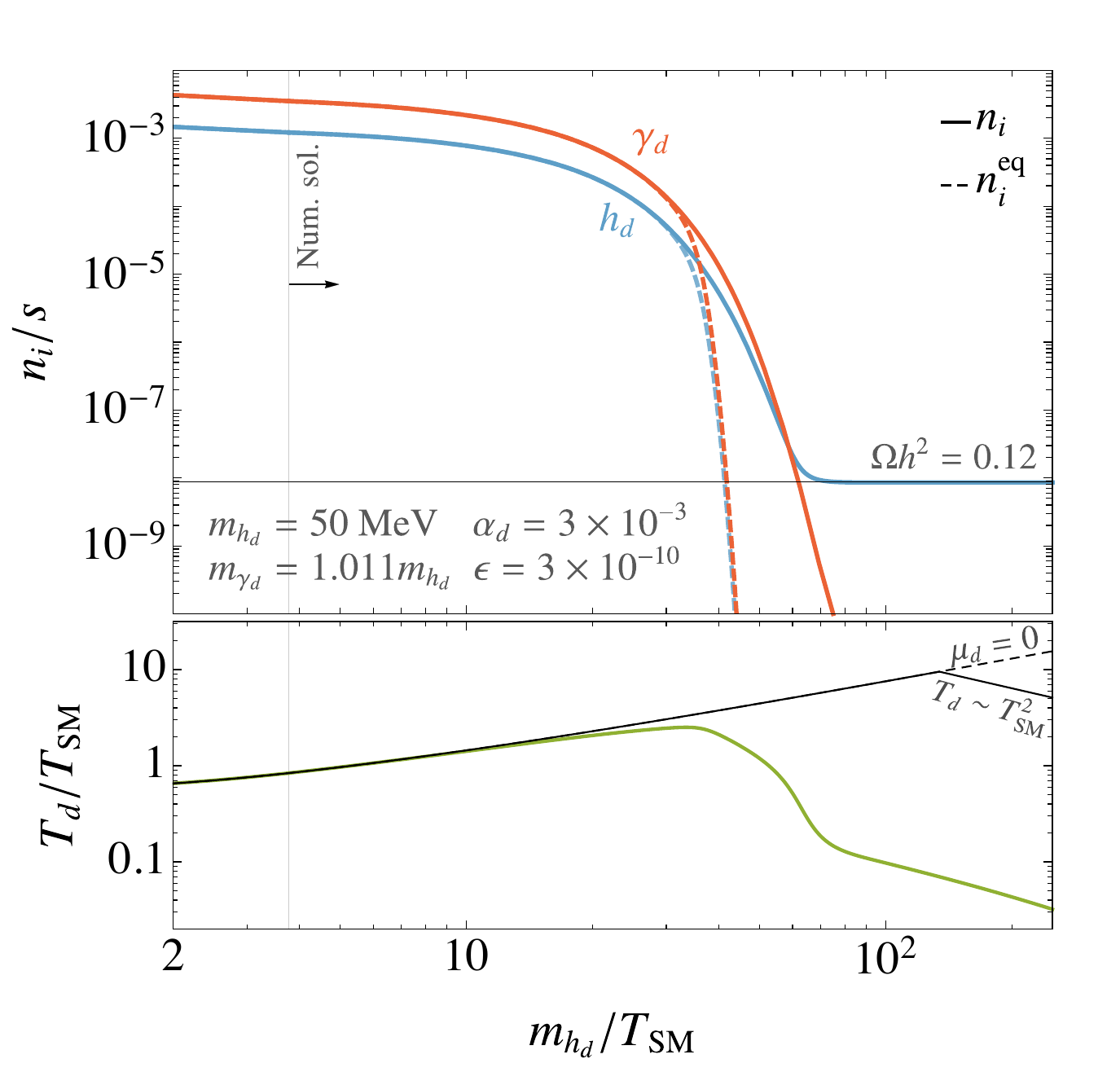}
  \caption{Same as Fig.~\ref{fig:FreezeIn_1d} for the secluded freeze-out regime. Separate entropy conservation in the two sectors is assumed to the left of the thin black vertical line, while the full set of BEs~(\ref{eq:boltz_nhd}-\ref{eq:boltz_Td}) are solved to the right. The evolution of $\Td$ for a completely decoupled DS with a sudden freeze-out of $3\rightarrow2$ processes at $\mhd/\TSM \approx 130$ is shown as a black curve in the bottom panel.}
  \label{fig:FreezeOut_1d}
\end{figure}

An example of the DS number densities and temperature evolutions showing the behavior described above is displayed in Fig.~\ref{fig:FreezeOut_1d}. We take $\TSM = \Td$ at 1 TeV and assume separate thermalization and entropy conservation in the two sectors until $\mhd/\TSM\approx4$ (vertical gray line in the figure). Below this temperature, the full set of BEs tracks the evolution of $\nhd$, $\nAd$, and $\Td$. In the bottom panel, the departure of the green curve from the black marks the onset of $\Ad$ decay. In the top panel, the departure of the solid curves from the dashed corresponds to the freeze-out of $3\lra 2$ processes, with the DS particles developing a non-zero chemical potential.

\begin{figure}
\centering
\includegraphics[width=0.45\textwidth]{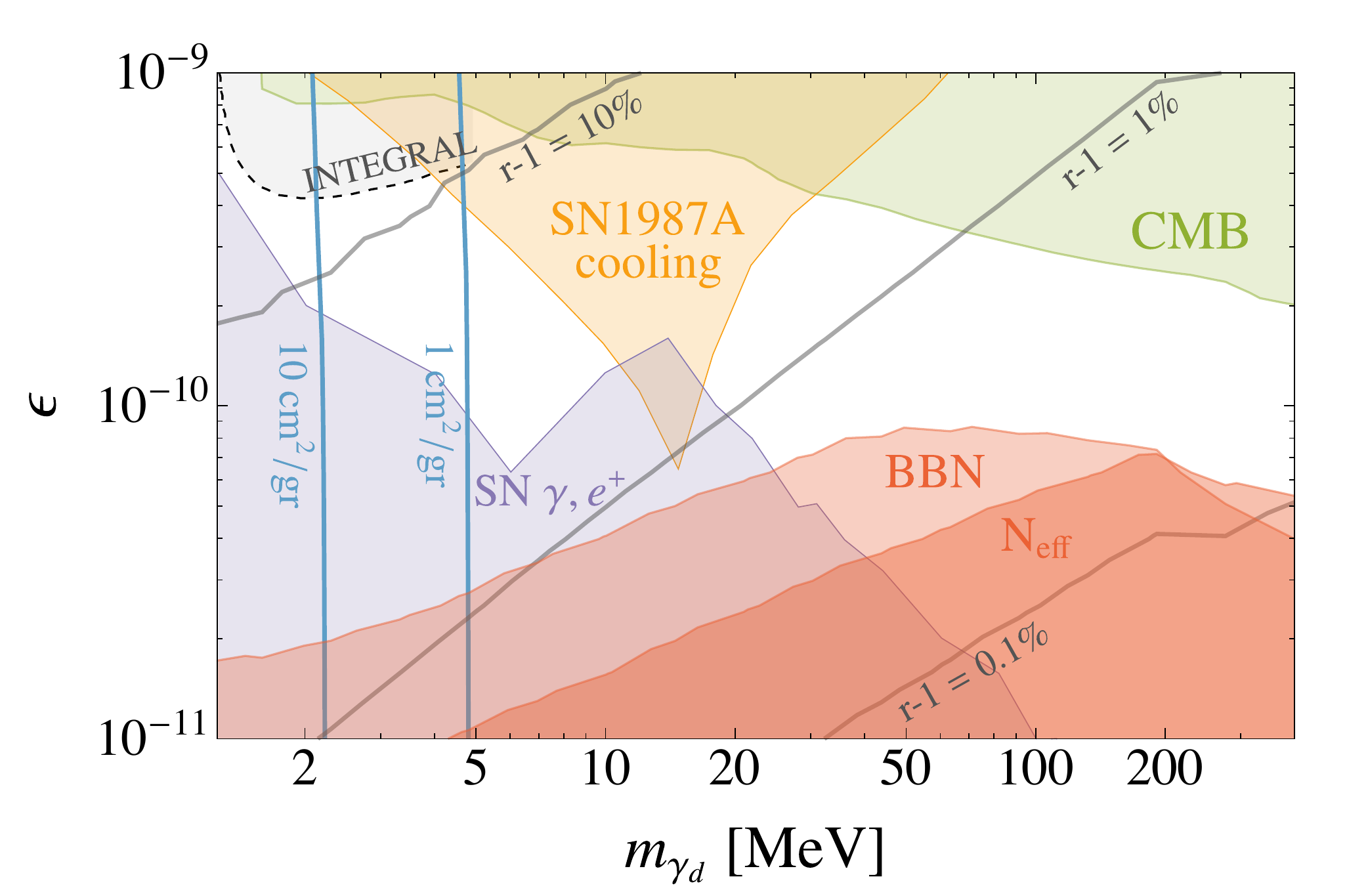}
\caption{Parameter space for the secluded annihilation freeze-out scenario for $\alpha_d = 3\times 10^{-3}$. The correct DHDM relic density is produced everywhere by fixing $r \equiv \mAd/\mhd$; constant values are shown as gray contours. Bounds include dark photon constraints from SN1987A cooling~\cite{Chang:2016ntp}, $e^+$ and $\gamma$ signals from $\Ad$ produced within SNe~\cite{DeRocco:2019njg}, CMB anisotropies limits on late DHDM decay~\cite{Slatyer:2016qyl}, and BBN and $N_\text{eff}$~\cite{Aghanim:2018eyx} bounds on the fraction of dark energy density. Also shown are the region of parameter space that could explain the SPI/INTEGRAL anomaly~\cite{Knodlseder:2003sv, Picciotto:2004rp}, and blue contours of constant $\sigma_{\rm SIDM}/\mhd$.}
\label{fig:FreezeOut_Pheno}
\end{figure}

\emph{Phenomenological Consequences.} The phenomenology of the secluded annihilation freeze-out scenario is summarized in Fig.~\ref{fig:FreezeOut_Pheno} for the choice $\alpha_d = 3\times 10^{-3}$. At every point, $r$ is chosen to satisfy $\Omega_\hd h^2 = 0.12$; gray contours correspond to constant $r$. Evidently, the spectrum becomes more compressed as $\mAd$ increases or $\epsilon$ decreases. Increasing $\alpha_d$ would lead to less degenerate masses.

Part of the parameter space is excluded by constraints on the dark photon. In the orange region, $\Ad$ induces excessive cooling of SN1987A~\cite{Chang:2016ntp} (this relies on the prevailing view that SN1987A was a core-collapse supernova; see however Ref.~\cite{Bar:2019ifz}). The purple region is excluded by the bound of Ref.~\cite{DeRocco:2019njg} described in the freeze-in phenomenology section.

Additional constraints are specific to the DHDM scenario. Electromagnetic energy injection due to dark Higgs decay at late times are constrained by measurements of CMB anisotropies by Planck~\cite{Ade:2015xua,Slatyer:2016qyl,Poulin:2016anj}, excluding the green region. The long-standing SPI/INTEGRAL observation of a positron excess in the galactic bulge~\cite{Knodlseder:2003sv} could be explained by decaying DM~\cite{Picciotto:2004rp, Hooper:2004qf}. Along the dashed black curve, DHDM can produce the observed photon flux, while in the gray region, too many $e^+$ are generated.

Small values of $\epsilon$ imply delayed $\Ad$ decays, leading to a prolonged period of cannibalism where the DS's energy density, $\rho_d$, dilutes less rapidly than the SM's, $\rho_\text{SM}$. A sizable fraction of $\rho_d$ during the time of BBN ($50\text{ keV} \lesssim \TSM\lesssim 1\text{ MeV}$) would increase Hubble and affect the observed primordial element abundances. We exclude parameters where $\rho_d > 0.1 \rho_\text{SM}$ at any time during BBN\@ (light red region). Moreover, if the additional $\rho_d$ is deposited into the photon bath through $\Ad$ decays after neutrino decoupling ($\TSM \simeq 2\text{ MeV}$), the photon temperature is increased relative to neutrinos, lowering the value of $N_\text{eff}$ measured by the CMB~\cite{Ibe:2019gpv}. The $95\%$ CL lower bound from Planck~\cite{Aghanim:2018eyx}, $N_\text{eff} > 2.55$, excludes the dark red region in the figure.  

Finally, the freeze-out scenario also allows for sizable self-interaction cross sections. The blue curves denote interesting values of $\sigma_{\rm SIDM}/\mhd$.

\emph{Conclusions.} 
This Letter presents a minimal realization of the $U(1)_d$ model that includes a DM candidate: a cosmologically stable dark Higgs. We identify three distinct production mechanisms, which are nontrivial realizations of either freeze-in or freeze-out from the primordial plasma. The parameter space, $10^{-13} \lesssim \epsilon \lesssim 10^{-6}$ and MeV $\lesssim \mAd \lesssim$ GeV, allows for viable DHDM with a rich cosmological history and diverse phenomenology.

\emph{Acknowledgement.} We thank H. Liu, P. Meade, A. Pierce, and H. Ramani for helpful discussions. 
JTR is supported by NSF CAREER grant PHY-1554858 and NSF grant PHY-1915409.
JTR acknowledges hospitality from the Aspen Center for Physics, which is supported by the NSF grant PHY-1607611.

\twocolumngrid

\bibliographystyle{apsrev}
\bibliography{DarkHiggsDMbib.bib}

\clearpage
\appendix
\onecolumngrid

\setcounter{equation}{0}
\setcounter{figure}{0}
\setcounter{table}{0}
\setcounter{section}{0}
\makeatletter
\newcommand{\dd}{{\rm d}}

\section{Boltzmann Equations for a Dark Sector}
\label{sec:BE_terms}

The microscopic evolution of an arbitrary DS particle's phase space distribution $f_\psid$ can be traced with the BE $\hat{L}[f_\psid] = C[f_\psid]$ \cite{Kolb:1990vq}, where $\hat{L}$ is the Liouville operator and $C[f_\psid]$ is the collision operator which involves particle species $\psi$. In a Friedmann-Robertson-Walker Universe the former reads $\hat{L}[f_\psid] \equiv E_\psid \dot{f}_\psid - H \bold{p}_\psid^2 \partial_{E_\psid} f_\psid$, where $E_\psid$ and $\bold{p}_\psid$ are the energy and momentum of $\psid$ particles. The $n$-th moment of the BE is
\beq
\frac{g_\psid}{(2\pi)^3} \int \frac{\dd\bold{p}_\psid}{E_\psid} E_\psid^n \hat{L}[f_\psid] = \frac{g_\psid}{(2\pi)^3} \int \frac{\dd\bold{p}_\psid}{E_\psi} E_\psid^n C[f_\psid] \, .
\eeq
The $0$-th and $1$-st moments can be written in terms of $\psid$ number density $n_\psid$, energy density $\rho_\psid$, and pressure $P_\psi$ as
\bea
\dot{n}_\psid + 3 H n_\psid & = & \frac{g_\psid}{(2\pi)^3} \int \frac{\dd\bold{p}_\psid}{E_\psi} C[f_\psid] \label{eq:BE_n} \, , \\
\dot{\rho}_\psid + 3 H (\rho_\psid + P_\psid) & = & \frac{g_\psid}{(2\pi)^3} \int \frac{\dd\bold{p}_\psid}{E_\psi}E_\psid C[f_\psid] \, . \label{eq:BE_rho}
\eea
Assuming that $\psid$ maintains kinetic equilibrium with the DS bath, the $1$-st moment of the BE tracks the evolution of the dark temperature, $\Td$. Note that deviations from equilibrium would require solving for higher moments of the BEs; any such higher order effects have been neglected in this study.

In the nonrelativistic regime, at leading order in $\Td/m_\psid$ and $p_\psid/m_\psid$ (where $p_\psid \equiv |\bold{p}_\psid |$ is the three-momentum), one can make the following replacements in Eq.~\eqref{eq:BE_rho},
\beq
E_\psid \approx m_\psid + \frac{\bold{p}_\psid^2}{2m_\psid}, \ \ \ \ \ \ \ \rho_\psid \approx n_\psid m_\psi \left(1 + \frac{3}{2} \frac{T_d}{m_\psid}\right), \ \ \ \ \ \ \ P_\psid \approx  n_\psid T_d,
\label{eq:rho_P_n}
\eeq
to obtain
\beq
\left(m_\psid + \frac{3}{2}T_d\right)\left(\dot{n}_\psid + 3 H n_\psid\right) + \frac{3}{2}n_\psid T_d \left(\frac{\dot{T}_d}{T_d} + 2 H\right) \approx \frac{g_\psid}{(2\pi)^3} \int \frac{\dd\bold{p}_\psid}{E_\psi}\left(m_\psid + \frac{\bold{p}_\psid^2}{2m_\psid}\right) C[f_\psid] \, .
\label{eq:BE_2nd_nr}
\eeq
Note that the 0-th order term in the nonrelativistic expansion cancels out by inserting Eq.~\eqref{eq:BE_n} into Eq.~\eqref{eq:BE_2nd_nr}. Explicitly writing the collision term for a general process, $j$, that involves $\psid$ particles, the result is,
\beq
\frac{3}{2}T_d\left(\dot{n}_\psid + 3 H n_\psid\right) + \frac{3}{2}n_\psid T_d \left(\frac{\dot{T}_d}{T_d} + 2 H\right) \approx - \int d\Pi_N (2\pi)^4 \delta^{4}\left(p_{\rm fin} - p_{\rm init}\right) |\mathcal{M}|_j^2 \left(\prod f_{\rm init} - \prod f_{\rm fin}\right)  \frac{\bold{p}_\psid^2}{2m_\psid} \, ,
\label{eq:BE_Tevol}
\eeq
where $p_{\rm init}$ and $p_{\rm fin}$ are the total initial and final state momenta, respectively, $d\Pi_N = \prod_i g_i/(2\pi)^3 \dd \bold{p}_i/(2 E_i)$ corresponds to the phase space integral over all incoming and outgoing particles, $|\mathcal{M}|_j^2$ is the matrix element squared averaged over initial and final spins (including symmetry factors for initial and final identical particles), and $\prod f_{\rm init}$ and $\prod f_{\rm fin}$ are products of the initial and final distribution functions, respectively. In this study, Maxwell-Boltzmann statistics is always assumed. In Eq.~\eqref{eq:BE_Tevol}, the terms proportional to $m_\psid$ have cancelled out. The collision term on the right hand side of the equation should include all energy exchanging processes involving $\psid$. This term has a particularly simple expression for decay processes of the form $\psid \rightarrow$ SM SM,
\beq
\dot{n}_\psid + 3 H n_\psid + n_\psid \left(\frac{\dot{T}_d}{T_d} + 2 H\right) = - \Gamma^0_{\psid} \left[ n_\psid - \frac{\TSM}{T_d} n_\psid^{\rm eq}(T) \right] \, .
\label{eq:BE_2nd_decay}
\eeq
Here, $\Gamma^0_\psid$ is the zero temperature decay width.

If all the particles in the DS are nonrelativistic and maintain kinetic equilibrium with each other throughout, summing over Eq.~\eqref{eq:BE_rho} (or equivalently Eq.~\eqref{eq:BE_2nd_nr}) for all dark $\psi$ particles, results in a collision term that includes only interactions between the DS and the SM\@. If these interactions can be neglected, the collision term vanishes and the total DS energy is conserved,
\beq
\sum_\psi \left[ \left(m_{\psid} + \frac{3}{2}T_d\right) \left(\dot{n}_{\psid} + 3 H n_{\psid} \right) + \frac{3}{2} T_d n_{\psid} \left(\frac{\dot{T}_d}{T_d} + 2 H\right) \right] = 0 \, ,
\label{eq:BE_2nd_sum}
\eeq
where, as stated, the sum is taken over all DS particles. From the above equation, the effect on the evolution of $T_d$ of mass conversion into kinetic energy, and vice versa, becomes evident by replacing $\dot{n}_{\psid} + 3 H n_{\psid}$ with the right hand side of Eq.~\eqref{eq:BE_n}.

For completeness, the full set of BEs for the model discussed in this study are given below. The $0$-th moments of the BEs for the two species in the DS including all relevant terms are
\bea
\dot{n}_{\hd} + 3Hn_{\hd} & = &  \mathcal{C}_{2\Ad \ra 2\hd}  - \mathcal{C}_{3\hd \ra 2\Ad} - \mathcal{C}_{3\hd \ra 2\hd} - \mathcal{C}_{2\hd\Ad \ra \hd \Ad} \, , \nn \\
\dot{n}_{\Ad} + 3Hn_{\Ad} & = & - \mathcal{C}_{\Ad \ra e^+e^-} - \mathcal{C}_{2\Ad \ra 2\hd} + \frac{2}{3} \mathcal{C}_{3\hd \ra 2\Ad} \, ,
\eea
with collision terms
\bea
\mathcal{C}_{\Ad \ra e^+e^-} & \equiv & \Gamma^0_{\Ad} \left[\frac{K_1(r x_d)}{K_2(r x_d)}\nAd - \frac{K_1(r x)}{K_2(r x)}\nAdeq(\TSM) \right] \, , \nn \\
\mathcal{C}_{2\Ad \ra 2\hd} & \equiv & \langle \sigma v \rangle_{\Ad\Ad \to \hd\hd} \left[ \nAdsq-\frac{{\nAdeq}^2}{{\nhdeq}^2} \nhdsq\right] \, , \nn \\
\mathcal{C}_{3\hd \ra 2\Ad} & \equiv & \frac{3}{3!2!}\langle \sigma v^2 \rangle_{\hd\hd\hd \to \Ad\Ad} \left[\nhdcu - \frac{{\nhdeq}^3}{{\nAdeq}^2} \nAdsq \right] \, , \nn \\
\mathcal{C}_{3\hd \ra 2\hd} & \equiv & \frac{1}{3!2!}\langle \sigma v^2 \rangle_{\hd\hd\hd \to \hd\hd} \left[\nhdcu - \nhdeq \nhdsq \right] \, , \\
\mathcal{C}_{2\hd\Ad \ra \hd\Ad} & \equiv &  \frac{1}{2!}\langle \sigma v^2 \rangle_{\hd\hd\Ad \to \hd\Ad} \left[\nhdsq \nAd - \nhdeq \nhd \nAd \right] \, . \nn \label{eq:Coll_Terms_1}
\eea
Here, $n_{i}^{\rm eq} \equiv n_{i}^{\rm eq}(T_d)$ is the equilibrium number density (with zero chemical potential) of particle $i=\hd, \Ad$ as a function of the hidden sector temperature, while $n_{i}^{\rm eq}(T_{\rm SM})$ is evaluated at the SM temperature. $K_{1/2}$ are modified Bessel functions of the 2-nd kind, $x_d \equiv \mhd/T_d$, $x \equiv \mhd/\TSM$, and $\langle \sigma v \rangle$ and $\langle \sigma v^2 \rangle$ denote the thermal averaged cross sections for $2 \rightarrow 2$ and $3 \rightarrow 2$ processes, respectively. The cross sections for $2 \rightarrow 3$ processes are obtained by computing $\sigma_{2 \ra 3}$ numerically with {\tt CalcHEP}~\cite{Belyaev:2012qa} and integrating over the center of mass energy to get the thermal average at different temperatures using the usual formula from Ref.~\cite{Gondolo:1990dk} adapted to $2 \ra 3$ processes. The expressions for the inverse processes are obtained by applying detailed balance. 

The dark temperature evolution equation in the nonrelativistic regime is
\bea
n \frac{\dot{T_d}}{T_d} + 2 H n & = & -\mathcal{C}^{(E)}_{\Ad \ra e^+e^-} + \mathcal{C}^{(E)}_{2\Ad \ra 2\hd} + \frac{2}{3}\mathcal{C}^{(E)}_{3\hd \ra 2\Ad} + \mathcal{C}^{(E)}_{3\hd \ra 2\hd} + \mathcal{C}^{(E)}_{2\hd\Ad \ra \hd\Ad} \, ,
\label{eq:App_Td_evol}
\eea
with collision terms
\bea
\label{eq:coll_E}
\mathcal{C}^{(E)}_{\Ad \ra e^+e^-} & \equiv & \Gamma^0_{\Ad} \left[\left(1-\frac{K_1(rx_d)}{K_2(rx_d)}\right)\nAd + \left(\frac{K_1(rx)}{K_2(rx)}-\frac{\TSM}{T_d}\right)\nAdeq(T) \right], \nn \\
\mathcal{C}^{(E)}_{2\Ad \ra 2\hd} & \equiv & \frac{2}{3}(r-1) x_d \mathcal{C}_{2\Ad \ra 2\hd} \, , \nn \\
\mathcal{C}^{(E)}_{3\hd \ra 2\Ad} & \equiv & \left[\left(1 - \frac{2}{3}r \right)x_d  + \frac{1}{2}\right] \mathcal{C}_{3\hd \ra 2\Ad}\, ,  \nn \\
\mathcal{C}^{(E)}_{3\hd \ra 2\hd} & \equiv & \left(\frac{2}{3} x_d + 1\right) \mathcal{C}_{3\hd \ra 2\hd} \, , \nn \\
\mathcal{C}^{(E)}_{2\hd\Ad \ra \hd\Ad} & \equiv & \left(\frac{2}{3} x_d + 1\right) \mathcal{C}_{2\hd\Ad \ra \hd\Ad} \, .
\eea
Note that Eq.~\eqref{eq:App_Td_evol} is only valid in the nonrelativistic regime and therefore one need only consider leading terms in the Bessel functions within $\mathcal{C}^{(E)}_{\Ad \ra e^+e^-}$. We have verified that the processes $\hd\Ad \lra e^+e^-$, $\Ad e^\pm \lra \gamma e^\pm$, and $\hd e^\pm \lra \Ad e^\pm$ are negligible. As expected for a nonrelativistic mediator, $2 \lra 2$ processes are subdominant with respect to $1 \lra 2$, due to the extra insertion of electromagnetic or dark gauge couplings. In the scenarios considered in this work, the Hubble parameter is always dominated by the SM energy density, but the contribution from the DS is included in the numerical results.

\section{Reproductive Freeze-In}
\label{sec:ReproductiveDM}

The 0-th moment of the BE, Eq.~\eqref{eq:BE_n}, can be written in the general form,
\beq
\dot{n}_\psi + 3 H n_\psi = \sum_j \frac{N_{\psi,j}}{S_j} \mathcal{C}_{{\rm init} \ra {\rm fin},j} \equiv \sum_j \frac{N_{\psi,j}}{S_j} n_{{\rm init},j}^{\rm eq} \langle \sigma v \rangle_{{{\rm init} \to {\rm fin}},j} \left( \frac{n_{{\rm init},j}}{n_{{\rm init},j}^{\rm eq}} - \frac{n_{{\rm fin},j}}{n_{{\rm fin},j}^{\rm eq}} \right) \, ,
\label{eq:BE_0_explicit}
\eeq
where the sum is over all relevant processes of the form ${\rm init} \ra {\rm fin}$, which change the number of $\psi$ particles (indexed by subscript $j$). $N_{\psi,j} \equiv N_{\psi,{\rm fin},j} - N_{\psi,{\rm init},j}$ is defined as the difference between the number of outgoing and incoming $\psi$ particles in process $j$. $S_j$ is a combinatorial factor which counts the number of ways to sort all initial and final particles in process $j$. Finally, $n_{{\rm init}/{\rm fin},j}^{({\rm eq})}$ are products of (equilibrium) number densities of all particles in the initial or final state of process $j$.

For the case where all particles in the interaction $j$ belong to the DS and are nonrelativistic, one can plug Eq.~\eqref{eq:BE_0_explicit} into Eq.~\eqref{eq:BE_2nd_sum}, which results in
\beq
\sum_\psi n_{\psi} \left(\frac{\dot{T}_d}{T_d} + 2 H\right) = - \sum_j \sum_\psi \frac{N_{\psi,j}}{S_j} \left( \frac{2}{3} \frac{m_{\psi}}{T_d} + 1 \right) \mathcal{C}_{{\rm init} \ra {\rm fin},j} \, .
\label{eq:BE_2nd_sum_rep}
\eeq

Consider a DS which consists of two species $A$ and $B$, as is the case in the current study. As described in the main text, the reproductive regime occurs when $2 \lra 2$ as well as $2 \ra 3$ interactions involving DS particles only are simultaneously faster than the Hubble rate. In this case, the $2 \lra 2$ interactions, $AA \lra BB$, $AA \lra AB$, and $BB \lra AB$ enforce chemical equilibrium between $A$ and $B$, namely
\beq
\frac{n_A}{\nAeq} \approx \frac{n_B}{\nBeq} \, .
\label{eq:chem_eq}
\eeq
The collision term for any $3 \lra 2$ interaction, $j$, within the DS can then be found by plugging Eq.~\eqref{eq:chem_eq} into the definition of $\mathcal{C}_{{\rm init} \ra {\rm fin},j}$ for such a process, Eq.~\eqref{eq:BE_0_explicit}. The result is
\bea
\mathcal{C}_{2 \ra 3,j} & = & n_{2,j}\langle \sigma v \rangle_{{2 \to 3},j} \left( \frac{n_{A/B}}{n_{A/B}^{\rm eq}} - 1 \right) \approx - n_{2,j} \langle \sigma v \rangle_{{2 \to 3},j} \, ,
\label{eq:C_reproductive}
\eea
where the second equality is valid for the case where $n_{A/B} \ll n_{A/B}^{\rm eq}$, as is typically the case for freeze-in. Thus, the reaction always proceeds only in the $2 \to 3$ direction until the inequality no longer holds, rapidly forcing the dark temperature to a value for which the chemical potentials for both $A$ and $B$ vanish, {\it i.e.}~$n_{A} \approx n_{A}^{\rm eq}$ and $n_{B} \approx n_{B}^{\rm eq}$. Importantly, this mechanism essentially wipes out all initial conditions of the dark temperature.

The rapid cooling described above can be seen explicitly by plugging Eq.~\eqref{eq:C_reproductive} into the temperature evolution Eq.~\eqref{eq:BE_2nd_sum_rep}, which gives
\bea
(n_A + n_B) \left(\frac{\dot{T}_d}{T_d} + 2 H\right) & = & \sum_j \left[ \frac{2}{3} \frac{M_{3,j} - M_{2,j}}{T_d} + 1 \right] \frac{\mathcal{C}_{2 \ra 3,j}}{S_j} \, ,
\label{eq:rapid_cooling}
\eea
where $M_{2,j}$ ($M_{3,j}$) is the sum of masses of incoming (outgoing) particles for the $2 \to 3$ interaction. Note that in the reproductive regime $\mathcal{C}_{2 \ra 3,j} < 0$, and therefore $M_{3,j} > M_{2,j}$ corresponds to a cooling term as should be expected, since this corresponds to a transfer of kinetic energy to rest mass. Even when $M_{2,j} = M_{3,j}$, the constant term within the parenthesis still enforces cooling. This is the result of removal of kinetic energy corresponding to the growth in number of particles in a $2 \to 3$ interaction. More generally, any number changing process, $j$, will induce a temperature drop when the inequality, 
\beq
\frac{M_{3,j} - M_{2,j}}{T_d} > -\frac{3}{2},
\eeq
is satisfied. 
Such cooling can be more rapid than Hubble cooling if,
\beq
\biggr|\frac{1}{S_j} \left[ \frac{2}{3} \frac{M_{3,j} - M_{2,j}}{T_d} + 1 \right] \frac{n_{2,j} \langle \sigma v \rangle_{{2 \to 3},j}}{n_A + n_B}\biggr| > 2H \, .
\eeq

\section{Freeze-out of Secluded Annihilations}
\label{sec:FO_analytic}
As described in the main text, the secluded annihilation freeze-out scenario considered in this study shares various features with previous studies in the literature. The general picture can be understood as follows.
\begin{itemize}
\item The small coupling to the SM does not allow for full thermalization between the two sectors and thus the entire process described below occurs with a nontrivial evolution of the dark temperature.
\item The DS features sizable $3 \lra 2$ interactions and thus, at intermediate temperatures, undergoes a period of cannibalism. When $T_d$ drops below min[$\mhd,\mAd$], this results in a dark temperature which scales logarithmically with the SM temperature.
\item The depletion of DS particles occurs through the decay of dark photons via $\Ad \to e^+e^-$, delaying the freeze-out of secluded annihilation, $\hd \hd \ra \Ad \Ad$, to $x \gg 20$. 
\item The mass hierarchy considered in this study is $\mhd<\mAd$, which corresponds to a DM annihilation process that is kinematically forbidden at zero temperature. This significantly alters freeze-out with respect to the cases of an inverted or degenerate mass spectrum. 
\end{itemize}
Clearly, the entire evolution is captured within the numerical solutions of the BEs (\ref{eq:boltz_nhd}-\ref{eq:boltz_Td}). However, much of the qualitative behavior of a DS in such a regime can be understood analytically. The goal of this appendix is to provide the relevant equations, together with an intuitive understanding of this evolution.

The initial conditions for the evolution are a DS which is completely decoupled from the SM, with some initial DS to SM entropy ratio, $\xi_s \equiv s_d/s_{SM}$.
Using the SM temperature, $\TSM$, as an independent variable, the dark temperature, $\Td(\TSM)$, can be obtained by requiring separate conservation of entropy in each sector. The low temperature behavior is particularly informative. When $\Td$ drops below min[$\mhd,\mAd$], if number changing processes remain active, the DS's chemical potential is forced to zero. For $\mhd/\Td \gg 5/2$, the total hidden sector entropy density is 
\beq
s_d \approx \frac{\mhd \nhdeq + \mAd \nAdeq}{\Td} \, .
\eeq
Using separate entropy conservation in each sector, one can directly solve for $x_d \equiv \mhd/\Td$ as a function of $x \equiv \mhd/\TSM$,
\beq
x_d \approx \ln \left(\frac{x^3}{1.7 \xi_s g_{*,s}}\right) \, ,
\label{eq:xd_Cannibal}
\eeq
which is the logarithmic dependence typical of a cannibalizing sector \cite{Carlson:1992fn, Pappadopulo:2016pkp}. Neglecting possible energy transfer to or from the SM, this equation describes the evolution of $\Td$ until number changing processes within the DS become slower than the Hubble rate. From that point onwards, the dark temperature evolves approximately as $x_d \propto x^2$. Note that, depending on the size of the kinetic mixing parameter, energy transfer between the DS and the SM via the process $\Ad \lra e^+e^-$ might not be completely negligible. However, the above result serves as a reasonable approximation in the limit of small portal coupling. 

Simultaneously, the annihilation process $\hd\hd \lra \Ad\Ad$, forces chemical equilibrium such that $\nhd/\nhdeq = \nAd/\nAdeq$, {\it i.e.}~the dark particles develop nonvanishing and equal chemical potentials. One can now use the requirement of chemical equilibrium together with the rough dark temperature evolution described above to approximately determine the SM (DS) temperature, $\xG$ ($\xG_d$), at which $\Ad$ begins decaying. This $\xG$ can be found by requiring that
\beq
\frac{\nAdeq(\xG_d)}{\nhdeq(\xG_d) + \nAdeq(\xG_d)} \Gamma^0_\Ad = H(\xG) \, .
\eeq

During the period where the decay process is active, and as long as $\hd\hd \lra \Ad\Ad$ exchanges remain efficient, depletion of the DHDM number density proceeds indirectly through the decay channel, since the $\Ad$ particles become exponentially less abundant as the temperature drops. To gain intuition regarding the evolution of the DS, it is insightful to solve an approximate BE in the limit where the two sectors are completely decoupled and cannibalism turns off abruptly at $\xG$. In this case, the temperatures of the two sectors relate via $x_d \approx \RG x^2$, where $\RG$ is set by the initial condition at approximately $\xG$. Then, as long as the inverse decay process can be neglected, the total number density of DS particles, $n = \nhd + \nAd$, can be approximated by
\beq
\dot{n} + 3Hn \approx -\frac{n^\eqt_\Ad}{n^\eqt} n \Gamma^0_{\Ad}  \approx \frac{n}{s} \frac{x}{1+\frac{1}{3r^{3/2}}e^{\Delta R_\Gamma x^2}}\frac{\Gamma^0_{\Ad}}{H_m} \, ,
\label{eq:BE_totn}
\eeq
where $r \equiv \mAd/\mhd$, $\Delta \equiv r-1$, and $H_m \equiv H(\mhd)$. Additionally, in the second equality the temperature dependence on the effective number of relativistic degrees of freedom has been neglected. Note that determining the temperature from which the evolution described by Eq.~\eqref{eq:BE_totn} holds, is nontrivial.

Within this approximation, Eq.~\eqref{eq:BE_totn} provides an analytic solution for the DHDM number density,
\beq
\label{eq:FI_sol_approx}
\frac{n}{s} = \mathcal{A} \left[1+3r^{3/2}e^{-\Delta R_\Gamma x^2}\right]^{\frac{\Gamma^0_{\Ad}}{2\Delta R_\Gamma H_m}-1} \, ,
\eeq
by also using the equality $\nhd/n = n^\eqt_\hd/n^\eqt$. The constant, $\mathcal{A}$, depends on the initial conditions at $\xG$. 
The above result is extremely informative. First, note that the total number of particles can only be reduced if $\Gamma^0_{\Ad}/2 H_m > 1$. Additionally, when $\Delta R_\Gamma x^2 \gg 1$, the yield, $n/s$, goes to a constant, {\it i.e.}~it becomes temperature independent. Thus, as long as the $\hd\hd \lra \Ad\Ad$ cross-section is large enough to ensure chemical equilibrium throughout the time where $\Ad$ decays are active, the final abundance does not depend on the magnitude of the cross-section for this process. Note that this behavior differs significantly from the one-way/co-decay and forbidden scenarios considered in the literature. Essentially, the constant yield is the result of a balance between chemical equilibrium in the DS and the effect of $\Ad$ decay. The specific requirements for this mechanism to occur are a decaying state that is heavier than the DM, and a DS temperature that scales as $x_d \propto x^2$. Further analysis and possible generalizations will be considered in a future study.

The numerical solution of the BEs in this regime can differ significantly from the approximation given by Eq.~\eqref{eq:FI_sol_approx}. This difference is mainly due to corrections to the scaling of the dark temperature, $T_d(T_{\rm SM})$ (see Fig.~\ref{fig:FreezeOut_1d}), and to the $T_{\rm SM}$ dependence on the effective number of degrees of freedom of the SM plasma; both of which were simplified in the derivation above. However the parametric dependence of this solution is reflected in the numerical results. Namely, the DHDM abundance is exponentially sensitive to the mass splitting $\Delta = r-1$, and to the kinetic mixing parameter $\epsilon$ (which controls the onset of $\Ad$ decay and therefore the value of $\xG$), while being almost independent of to the $\hd\hd \lra \Ad\Ad$ cross-section.

\end{document}